\documentstyle[here,epsfig]{mn2e}
\begin{document}
\def\simlt{\mathrel{\rlap{\lower 3pt\hbox{$\sim$}}\raise 2.0pt\hbox{$<$}}}
\def\simgt{\mathrel{\rlap{\lower 3pt\hbox{$\sim$}} \raise
2.0pt\hbox{$>$}}}

\newcommand{\lsim}{\raisebox{-0.13cm}{~\shortstack{$<$ \\[-0.07cm] $\sim$}}~}
\newcommand{\gsim}{\raisebox{-0.13cm}{~\shortstack{$>$ \\[-0.07cm] $\sim$}}~}

\def\whs{\thinspace ${\rm W \: Hz^{-1} \: sr^{-1}}$}
\def\lpow{\thinspace ${\rm log_{10}P_{1.4GHz}\: }$}
\def\kv{\thinspace $K_{Vega}$}
\def\kab{\thinspace $K_{\rm AB}$}

\title[] 
{A new measurement of the evolving near-infrared 
galaxy luminosity function out to $z \simeq 4$: a continuing challenge to 
theoretical models of galaxy formation} 

\author[M. Cirasuolo et al.] {
\parbox[h]{\textwidth}{M. Cirasuolo$^{1}$, R. J. McLure$^{1}$, J. S. Dunlop$^{1,2}$, 
O. Almaini$^{3}$, S. Foucaud$^{3}$,  C. Simpson$^{4}$.\\ } 
\vspace*{2pt} \\
$^{1}$SUPA\thanks{Scottish Universities Physics Alliance} Institute
for Astronomy, University of Edinburgh, 
Royal Observatory, Edinburgh EH9 3HJ\\
$^{2}$Department of  Physics and  Astronomy, University of British Columbia, 
6224 Agricultural Rd., Vancouver, B.C., V6T 1Z1, Canada \\
$^{3}$School of Physics and Astronomy, University of Nottingham,
University Park, Nottingham NG7 2RD\\ 
$^{4}$Astrophysics Research Institute, Liverpool John Moores
University, Twelve Quays House, Egerton Wharf, Birkenhead CH41 1LD}
\maketitle 
\begin{abstract}
We present the most accurate measurement to date of cosmological evolution of the near-infrared galaxy
luminosity function, from the local Universe out to $z \simeq 4$. The analysis is based on a large
and highly complete sample of galaxies selected from the first data release of the UKIDSS Ultra
Deep Survey. Exploiting a master catalogue of $K$- and $z$-band selected galaxies 
over an area of 0.7 square degrees, we analyse a
sample of $\simeq 50,000$ galaxies,
all with reliable photometry in 16-bands from the far-ultraviolet to the mid-infrared.
The unique
combination of large area and depth provided by 
the Ultra Deep Survey allows us to trace the evolution of the $K$-band luminosity function with
unprecedented accuracy. In particular, via a maximum likelihood analysis we obtain 
a simple parameterization for the luminosity function and its cosmological evolution, including both luminosity
and density evolution, which provides an excellent description of the data from $z =0$ up to $z \simeq 4$.
We find differential evolution for galaxies dependent on galaxy luminosity, 
revealing once again the ``down-sizing behaviour'' of galaxy formation.
Finally, we compare our results with the predictions of the latest theoretical models of galaxy
formation, based both on semi-analytical prescriptions, and on full hydrodynamical simulations.

\end{abstract}
\begin{keywords} galaxies: evolution - galaxies: formation - cosmology:
observations
\end{keywords}
%
\section{INTRODUCTION}
In recent years outstanding progress has been made in understanding the structure and
the evolution of the Universe. In particular, the canonical Lambda cold dark matter ($\Lambda$CDM) cosmological
model now provides a solid framework and is able to explain a large variety of
observations: from the fluctuations in the cosmic microwave background radiation (CMB) 
to the  large scale structure of the Universe (e.g. Perlmutter et al. 1998; Riess et al. 1998; 
Tegmark et al. 2004; Spergel et al. 2003, 2007). Moreover, the $\Lambda$CDM model also provides a 
clear picture for the composition of the Universe, with baryonic matter 
contributing only $\simeq 4$\% to the density and the Universe being dominated by dark energy 
and dark matter. Over the last fifteen years, enormous improvements in numerical N-body 
simulations have allowed the hierarchical growth of the dark matter (DM) to be accurately traced, 
from the initial perturbations to the large scale structure we see in the local Universe (e.g. see Springel et al. 2005).
However, while the evolution of the DM is now well understood, understanding
the physics involved in the evolution of the baryons is much more complex. While the 
evolution of the DM is only subject to gravity, many other physical processes must be taken into account to 
describe the formation and evolution of galaxies: e.g.  gas cooling, star-formation, 
production of dust and metals and feedback from supernovae and nuclear black holes.

Various theoretical models have been developed either by implementing  semi-analytical 
prescriptions for the transformation of gas into stars on the DM merger tree 
(e.g. Kauffmann et al. 1993; Cole et al. 2000, Somerville et al. 2001) or by solving a full set 
of hydrodynamical equations for the DM and baryons with smooth particle hydrodynamics (SPH) 
simulations (e.g. Nagamine et al. 2000, 2001; Springel \& Hernquist 2003). 
Over the last decade, these theoretical models have become increasingly successful 
in some regards in explaining the observed properties of galaxies in the local Universe. 
Traditionally, however, most semi-analytical models have demonstrated a 
tendency to predict very few old/massive galaxies at high redshift ($z \simgt 1$) because, in the hierarchical growth of 
DM, 
large structures naturally form late by continuous  merging  of smaller haloes 
(e.g. Cole et al. 2000; Menci et al. 2002). However, this hierarchical behaviour seems 
to be in contrast with recent observational results, with the advent of recent near-infrared surveys, in particular, 
revealing 
the presence of a substantial population of old, massive evolved galaxies (Extremely Red Objects, Distant Red Galaxies) at 
$z \simgt 1.5$, which were previously missed by optical surveys (Cimatti et al. 2002; Daddi et al. 2002).

Several deep multi-wavelengths surveys are now ongoing or have just been completed 
(Cimatti et al. 2002;  Giavalisco et al. 2004; 
Rix et al. 2004; Lawrence et al. 2007; Scoville et al. 2007) with the aim of tracing the 
assembly history of galaxies. Exploiting the near-infrared data 
and supported by spectroscopic and photometric redshift estimations, several studies have attempted to 
derive the evolution of the near-infrared luminosity 
function (LF), as a tracer of the mass assembly history of galaxies. 
In fact, it has long been understood that observations of the rest-frame near-infrared emission 
of galaxies is the best tracer of the stellar mass because it is less affected by dust absorption
and ongoing star-formation (e.g. Lilly \& Longair 1984; Dunlop et al. 1989; 
Glazebrook et al. 1995; Cowie et al. 1996; Poggianti 1997 ). 

Several authors have found evidence for evolution of the 
near-infrared LF with redshift, with the evolution being very slow from the local Universe up to 
$z \sim 0.5$ and then progressively faster (Glazebrook et al. 1995; Bolzonella et al. 2002; 
Pozzetti et al. 2003;  Dickinson et al. 2003; Drory et al. 2003; Caputi et al. 2006; 
Saracco et al. 2006). However, as shown by Saracco et al., there are still
some discrepancies between results obtained by different authors, especially at the bright end, 
presumably due to the small-number statistics imposed by the limited areal coverage of the available data.
Recently, the advent of the Spitzer satellite has provided the opportunity to derive reliable 
estimates for the stellar masses in galaxies at high redshift, allowing a direct 
study of the evolution of the mass function (Fontana 2004, 2006; Drory et al. 2005; 
Perez-Gonzalez et al. 2008; Pozzetti et al. 2007). 
These studies have consolidated the ``downsizing'' scenario (Cowie et al. 1996) in which 
the more massive galaxies form their stars at higher redshift and on a shorter time scale
compared to less massive systems.

From the theoretical point of view, this apparent discrepancy between the hierarchical growth 
of DM and the downsizing evolution observed in galaxies seems to be reconciled in the latest
models by properly taking into account both stellar and nuclear feedback 
(Granato et al. 2001, 2004; Cirasuolo et al. 2005; Di Matteo et al. 2005; Croton et al. 2006;
Bower et al. 2006; Menci et al 2006; Monaco et al. 2007).

The aim of this work is to obtain an extremely accurate determination of the 
cosmological evolution of the near-infrared galaxy LF which can then 
be wielded as a powerful tool to constrain and possibly discriminate 
between the various prescriptions for the physical mechanisms implemented in galaxy-formation 
models.

The layout of the paper is as follows.
In Section 2 we summarize the main properties of the datasets 
used in this work and in Section 3 we 
describe in detail  the method 
we have developed for the estimation of redshifts from the 
available multi-band photometry. 
In Section 4 we derive the near-infrared LF and its cosmological evolution, comparing it 
with previous results found in literature. Finally in Section 5 we compare our results with the
predictions of some of the latest models of galaxy formation. 
Our conclusions are presented in Section 6.

Throughout this paper all magnitudes are quoted in the AB system 
(Oke \& Gunn 1983) and we have adopted a concordance cosmology 
with $\Omega_{\rm M}=0.3$, 
$\Omega_\Lambda=0.7$ and $H_0=70~{\rm km~s^{-1}}$.

\section{The Data-sets}
The aim of this paper is to analyse the evolutionary properties of galaxies
by exploiting the Ultra Deep Survey (UDS),  the deepest of the five surveys that constitute
the UKIRT Infrared Deep Sky Survey (UKIDSS - Lawrence et al. 
2007). For the purpose of this work we used data from the  first data release of the survey
(DR1; Warren et al. 2007), which includes imaging in the $J$- and $K$-bands with $5 \sigma$ depths of 
23.3 and 23.1 respectively within a $3$-arcsec diameter aperture. 
Crucially, the UDS field benefits from a large area (0.8 square degrees) 
and deep multi-wavelength coverage. 
For this work we primarily exploit the deep optical data from the Subaru 
XMM-Newton Deep Survey (SXDS; Sekiguchi et al. 2005; Furusawa et al. 2008), 
which covers most of  the UDS field  in  $BVRi'z'$ filters to typical 
$5 \sigma$ depths of $B=27.5$, $V=26.7$, $R=27.0$, $i'=26.8$ and $z'=25.9$ 
(within a $3$-arcsec diameter aperture).

\begin{figure}
\center{{
\epsfig{figure=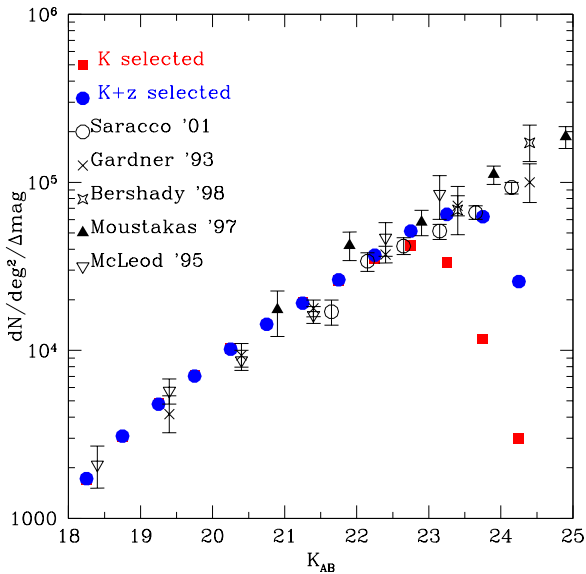, height=8cm}
\caption{\label{counts} Number counts in the $K$-band for sources in the UDS field. The $K$-band 
selected (solid squares) and $K+z'$ selected (filled dots) samples are described in the text. 
For comparison we also show the number counts in the $K$-band from Saracco et al. 
(2001, open circles), Gardner et al. (1993, crosses), Bershady et al. (1998, stars) Moustakas
et al. (1997, filled triangles), and McLeod et al. (1995, open triangles).}
}}
\end{figure}

For the construction of the initial galaxy catalogue we performed source extraction on
the $K$-band mosaic by using the public code {\sc sextractor} (Bertin \& Arnout
1996). The number counts for the $K$-band selected sources are shown in Fig.
\ref{counts} and compared with previous estimates found in the literature. 
In order to improve the completeness of our sample at faint
magnitudes, we also performed source extraction on the $z'$-band images from
the SXDS. The advantage of doing this is that the $z'$-band images are $\sim 2.5$
magnitudes deeper than the DR1 K-band data in the UDS. 
Therefore, performing aperture photometry on the K-band images centered at the position 
of the $z'$-band extracted sources allowed us to
increase the completeness close to the $K$-band flux limit, at least for sources 
which do not have extremely red colours (e.g. $z'-K  < 2$).

This technique allowed us to build a {\it master-catalogue} including 
all the sources extracted directly from the $K$-band images, plus all the sources
selected in the $z'$-band  with a counterpart in the $K$-band mosaic. 
The number counts for sources in the master-catalogue ($K+z'$ selected) are shown in
Fig. \ref{counts} with filled dots. As can be seen from Fig. 1, the procedure of
selecting sources in both the $K$- and $z'$-bands has increased the completeness of 
the sample to $\simeq 100$\% at $K\leq23$.

To fully exploit the high completeness of our sample, for the purpose of this work we only 
considered sources in the master-catalogue with $K \le 23$ and limit the analysis to 
an area of 0.7 square degrees which is covered by both the UDS and SXDS imaging. 
In order to model the spectral energy
distribution (SED) of these sources, we performed aperture photometry on the 
$BVRi'z'JK$ images centered at the position of the $K+z'$ selected sources. The
optical and near-infrared photometry were measured in 3$^{\prime \prime}$
diameter apertures, and then point spread function (PSF) corrected to 
total magnitudes. The consistent image quality of the optical and near-infrared data,
with Full Width Half Maximum (FWHM) in the range $0.75 <
\rm FWHM < 0.80$ arcsec, means that the differential aperture corrections between
bands are small ($\le 0.1$ mag). 

The UDS field also benefits from optical data obtained with the 
Canada-France-Hawaii telescope (CFHT) as part of the CFHT Legacy Survey. 
Source catalogues in $u,g,r,i,z$  bands from the TERAPIX release 3 (T0003) were retrieved from the CHFT legacy survey 
website{\footnote{http://www.cfht.hawaii.edu/Science/CFHLS/}} and 
matched with the sources in the master catalogue by using a searching radius 
of 1 arcsec. The typical $5 \sigma$ depths for the CFHT data ({\sc mag\_auto})
are: $u=25.3$, $g=25.2$, $r=24.7$, $i=24.4$ and $z=22.5$. 

Extremely valuable coverage of the UDS at mid-infrared wavelengths 
is provided by the  {\it Spitzer} satellite as
part of the {\it Spitzer} wide-area infrared extragalactic (SWIRE) survey 
(Lonsdale et al. 2003; 2004). 
For the present study we exploited the IRAC data at 
$\rm 3.6 \mu m$ and  $\rm  4.5 \mu m$ presented in the second data release  
(Surace et al. 2005). We performed aperture photometry on the IRAC images 
centered at the position of the $K+z'$-selected sources within a 
3$^{\prime \prime}$-diameter aperture. Since the mean FWHM of the IRAC PSF 
at $\rm 3.6 \mu m$ is 1.66 arcsec (Fazio et al. 2004) the magnitudes required
aperture correcting to match the total magnitudes. We performed tests studying
the curve of growth of isolated stars in the field and derived
aperture corrections of 0.53 and 0.58 magnitudes for the 
$\rm 3.6 \mu m$ and  $\rm  4.5 \mu m$ channels respectively, in agreement with 
Surace et al. (2005).

At shorter wavelengths the available multi-wavelength data in the UDS has recently been complemented 
by observations in the ultraviolet obtained with the {\it GALEX} satellite (both FUV and NUV channels). 
Given the large FWHM of the  {\it GALEX} PSF (6$^{\prime \prime}$) we used the total magnitudes given in the {\it GALEX} 
archive catalogues{\footnote{http://galex.stsci.edu/GR1/}}  and matched sources by using a searching radius of 2 arcsec.

Star-galaxy separation has been performed by combining information from 
the stellarity parameter, as measured by {\it SExtractor} on the SXDS images, with 
the position of the sources in the $BzK$ colour-colour diagram
(Daddi et al. 2004). As pointed out by Daddi et al., 
stars have colours that are
clearly separated from the region occupied by galaxies and can be efficiently
isolated with the criterion ${\rm (z' - K) < 0.3 (B - z') - 0.5}$.
It is worth noting that the vast majority
of these sources classified as stars also show an unacceptable value of the $\chi^2$ when 
fitted  with galaxy templates.

To summarise, the master-catalogue of sources we used for this work has been
selected by using both the UKIRT WFCAM $K$-band and Subaru SuprimeCam $z'$-band images in the UDS field.
For the following analysis we consider $\simeq$50,000 sources with $K \le 23$ over an
area of 0.7 square degrees, with each source having reliable
photometry (detections or upper-limits) in 16 broad-bands 
from the far-ultraviolet to $\rm  4.5 \mu m$.

\section{Redshift determination}
The extensive and deep multi-wavelength information in the UDS field is
vital for detailed modelling of the galaxy spectral energy distributions (SEDs). Following Cirasuolo et al.
(2007), we derived photometric redshifts for all the sources in the $K+z'$ 
master catalogue by fitting the observed photometry with both empirical and
synthetic galaxy templates.

\subsection{Photometric redshifts}
The fitting procedure to derive photometric redshifts, 
based on $\chi^2$ minimization, was performed
with a code based largely on the public package {\it Hyperz} (Bolzonella,
Miralles \& Pell\'{o} 2000). As empirical templates we used both the average
local galaxies SEDs derived by Coleman, Wu and Weedman (CWW) and
the more recent templates obtained within the K20 survey (Cimatti et al. 2002; 
Mignoli et al. 2005). In order to improve the characterization of young, blue
galaxies we also implemented six SEDs of observed starbursts from Kinney et al. 
(1996). 

To generate synthetic templates of galaxies we used the stellar population 
synthesis models of Bruzual \& Charlot (2003)
assuming a Salpeter initial mass function (IMF) with a lower and upper
mass cutoff of 0.1 and  100 ${\rm M_{\odot}}$ respectively.  
We used a variety of star-formation histories;
instantaneous-burst and exponentially-declining star formation with $e$-folding
times $\rm 0.3 \leq \tau (Gyr)\leq 15$, all with a fixed solar metallicity.
Dust reddening was taken into account by 
following the obscuration law 
of Calzetti et al. (2000) within the range $0 \leq A_V \le 2$. We also added a
prescription for the Lyman series absorption due to the HI clouds in the inter
galactic medium, according to Madau (1995). Finally, at each redshift, we only
permitted models with an age less than the age of the Universe at that redshift.

\subsection{Spectroscopic redshifts}
\begin{figure}
\center{{
\epsfig{figure=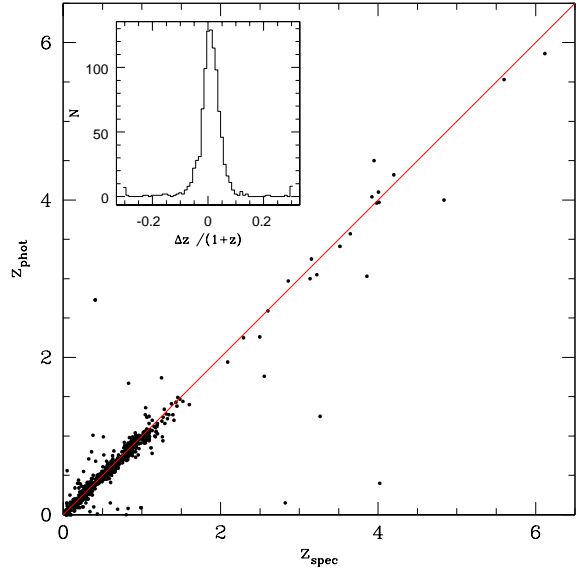, height=8cm}
\caption{\label{zz} Photometric redshift plotted versus 
spectroscopic redshift for $\sim 1200$ galaxies in the UDS sample with secure spectroscopic redshifts 
(Yamada et al. 2005; Akiyama et al. in preparation; 
Smail et al. in preparation; Simpson et al. 2006a; Simpson et al. in 
preparation).  
The small inset shows the
distribution of $\Delta z/(1+z)$.
The agreement is remarkably good; the mean value of the distribution of 
$\Delta z/(1+z)$ is 0.008, with a standard deviation $\sigma =0.034$,  
excluding the clear outliers (see text for details).}
}}
\end{figure}

The UDS also benefits from various large spectroscopic campaigns that allow us to test
the reliability of our redshift estimations. Fig. \ref{zz}  shows the
comparison of our photometric redshifts estimates for $\sim 1200$ galaxies in
the UDS field with secure spectroscopic redshifts (Yamada et al. 2005; 
Akiyama et al. in preparation; Smail et al. in preparation; Simpson et al. 
2006a).
The agreement is remarkably good over the full redshift range $0 <z <6$. The
fraction of outliers (defined as sources with 
$|\Delta z|/(1+z) \equiv  |(z_{\rm spect} - z_{\rm phot}) |/ 
(1+z_{\rm spect})  > 0.15$) is extremely low, less than 2\%. The improvement
compared with our previous work (Cirasuolo et al. 2007) is mainly due to the
added value of the $u$-band and Galex data, which significantly helps to remove the
degeneracy between double minima in redshift space.

More quantitatively, the accuracy of our photometric redshifts can be estimated
by looking at the distribution of the $\Delta z/(1+z)$ as shown in the small
inset in Fig. \ref{zz}.  The distribution has a mean consistent with zero
(0.008) and a standard deviation  $\sigma = 0.034$ (excluding the few
outliers with $|\Delta z|/(1+z) > 0.15$). This accuracy
is comparable to the best available from other surveys such as GOODS and COSMOS
(Caputi et al. 2006; Grazian et al. 2006; Mobasher et al. 2007). As already
pointed out in our previous work (Cirasuolo et al. 2007), this accuracy is preserved down to
faint $K$-band magnitudes ($K \ge 22$) and up to very high redshifts ($z > 5$). The ongoing
large spectroscopic campaign (PI Almaini)  with the Very Large Telescope 
to observe galaxies in the UDS with both VIMOS and FORS2 will, in the near
future, allow us to further test and refine our results.

\section{Luminosity Function}\label{lfsect}

Building on the reliable SED fitting procedure described previously, in this Section we 
exploit our large sample of UDS galaxies to characterise the
evolution of the rest-frame near-infrared LF.
Following Cirasuolo et al. (2007), the rest-frame absolute $K$-band magnitudes 
have been computed by using the closest observed band to the rest-frame $K$-band
depending on the redshift of the source. 
This method in fact minimises the uncertainties related to the  
$k$-corrections. 

\begin{figure*}
\center{{
\epsfig{figure=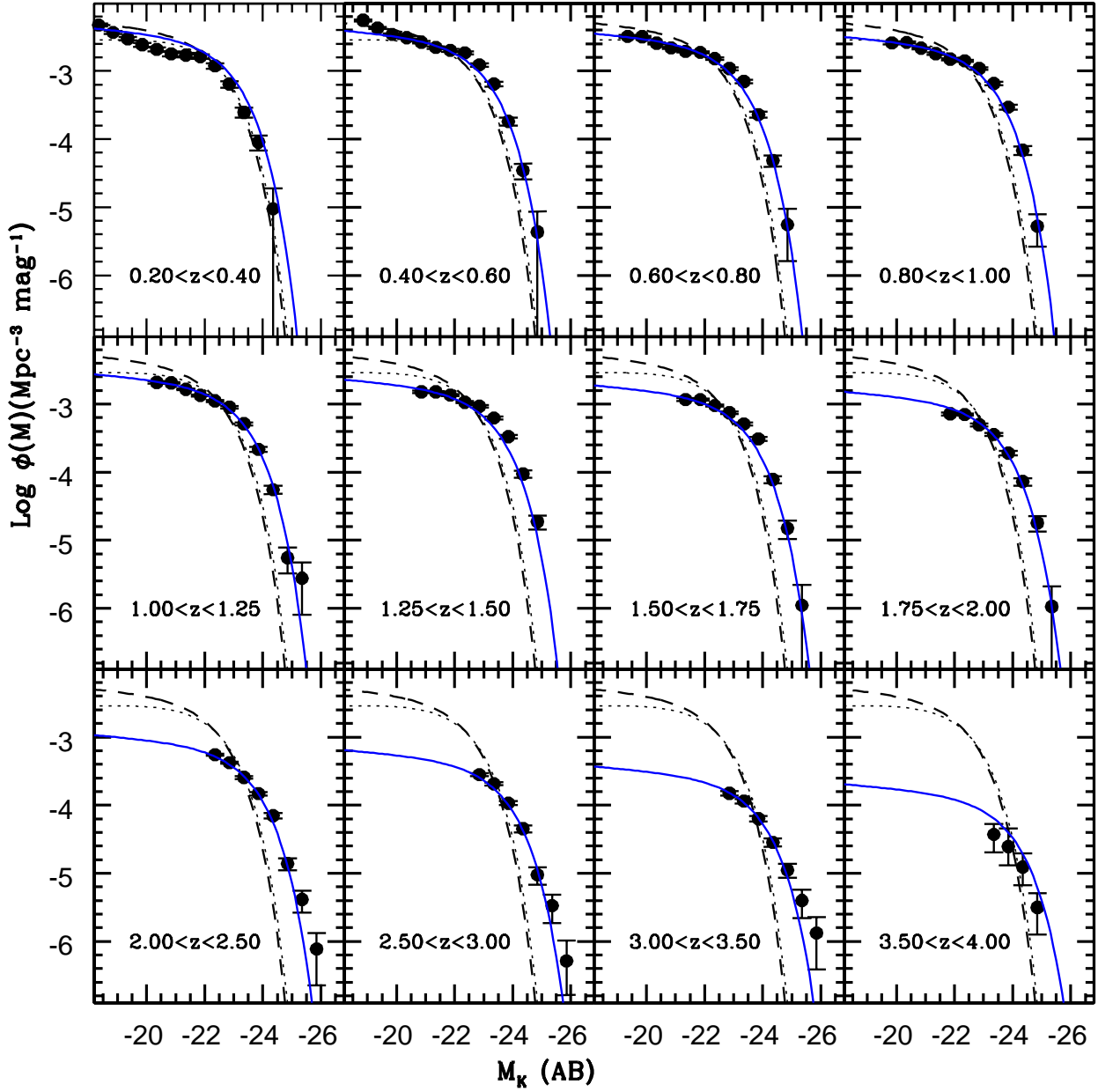, height=18cm}
\caption{\label{lf} Rest-frame $K$-band luminosity function in twelve redshift bins in the range
$0.2 \le z \le 4$. The solid dots represent the LF obtained with the $ 1/V_{max}$ method 
for sources in the UDS sample, while the solid line is the
best fit Schechter function obtained from the likelihood analysis and plotted at the mean
redshift of each bin. For comparison, the dashed and dotted lines 
are the LFs at $z=0$ obtained by Kochanek et al. (2001) and Cole et al. (2001), respectively.}
}}
\end{figure*}

The rest-frame $K$-band LF in the redshift range $0.2 \simlt z \simlt 4$ 
has been computed by using two independent methods: the 
$1/V_{max}$ method (Schmidt 1968) and the maximum likelihood
analysis of Marshall et al (1983). Our results are shown in Fig. \ref{lf}.
The LF in twelve redshift intervals computed via $1/V_{max}$ method is shown
in the figure as solid dots. The error-bars on the determination of the space
density of sources in each redshift and magnitude bin account for both
Poissonian statistical errors and the uncertainty in the 
photometric redshifts. The latter has been estimated by reconstructing several 
LFs, perturbing the value of the photometric redshift within its $1 \sigma$
confidence interval. For each source the SED-fitting procedure provides the 
best-fit redshift (identified by the lowest value of the $\chi^2$) and the
$1 \sigma$ interval defined as the redshift space around the best-fit minimum 
with $\Delta \chi^2 \equiv \chi^2_z - \chi^2_{min}\le 1$ 
(marginalising over the other free parameters in the
SED fitting e.g. age, reddening etc.). We performed Monte Carlo realizations, 
with each source being allocated a random value 
for its redshift within its $1 \sigma$ confidence interval, and having its luminosity recomputed appropriately before 
reconstructing the LF. The errors reported in Fig. \ref{lf} reflect the maximum differences in the LF
obtained from these different realizations.

An independent estimation of the LF has been obtained by using the maximum likelihood 
analysis. This method is less biased by strong 
density inhomogeneity  in the field, compared to the $1/V_{max}$ method, but
requires an {\it a priori} assumption regarding the shape of the LF. This latter has
been parameterised with a Schechter function (Schechter 1976): 
\begin{equation}
\phi(M) = 0.4 {\rm ln(10)} \phi_0 10^{-0.4 \Delta M (\alpha +1)} \rm exp(-10^{-0.4\Delta M})
\end{equation} 
with $ \Delta M = M_K - M^*_K$ assuming both a luminosity and density evolution
with redshift parameterised as:
\begin{equation}
M^*_K (z)  =  M^*_K (z=0) - \left(\frac{z}{z_M}\right)^{k_M},
\end{equation}
\begin{equation}
\phi_0(z)  =  \phi_0(z=0) \times exp\left[-\left(\frac{z}{z_{\phi}}\right)^{k_\phi} \right]
\end{equation}
The local value of the characteristic luminosity, $M^*_K (z=0)$, has been fixed to match the 
local value $-22.26 \pm 0.05$ (converted to AB using $K_{AB} = K_{Vega} +1.9$
and using $h=0.7$) derived by  Kochanek et al. (2001) from 
the Two Micron All Sky Survey (2MASS). 
It is worth noting that this value is very close to the value 
$-22.31 \pm 0.03$ derived by Cole et al. (2001).  
The maximum likelihood analysis was
performed by minimizing over the other five free parameters: the faint end 
slope $\alpha$ and four parameters ($z_M, k_M, z_{\phi}, k_\phi $) describing the evolution with
redshift. The best fit values for these parameters are given in Table \ref{tab_lf}, along with
the overall normalization at redshift zero, $\phi_0(z=0)$, obtained {\it a posteriori} by fitting 
the observed number counts. It is reassuring that the value of $\phi_0(z=0)$ we obtain is in
excellent agreement with the local value of $0.0039 \; \rm Mpc^{-3} $  and 
$0.0037 \; \rm Mpc^{-3}$, derived by Kochanek et al (2001) 
and Cole et al. (2001), respectively.

\begin{figure*}
\center{{
\epsfig{figure=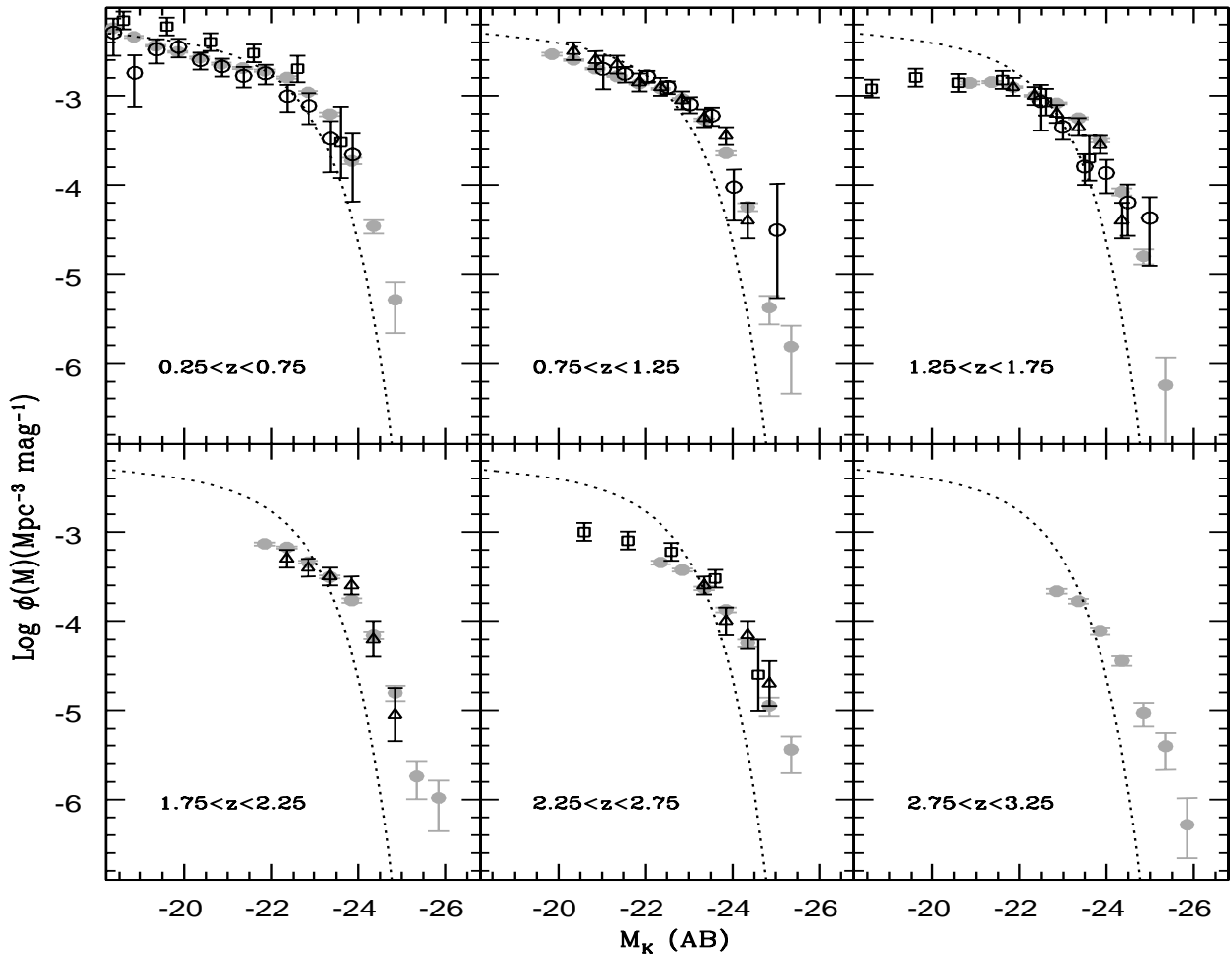, height=14cm, width= 18cm}
\caption{\label{lf_data} Comparison of our estimation of the $K$-band LF (solid gray dots) 
with previous results found in the literature. Open circles show the LF obtained by Pozzetti et al.
(2003) from the K20 survey, open triangles are the Caputi et al. (2006) results from the
GOODS/CDFS and open squares illustrate the estimates obtained by Saracco et al. (2006) from the
HDF-S. The dotted line is the local luminosity function of Kochaneck et al. (2001).}
}}
\end{figure*}

\begin{table}
\begin{center}
\caption{\label{tab_lf} Best fit parameters for the Schechter LF and its redshift evolution.}
\begin{tabular}{ll} \hline \hline
$\alpha$  &  $\phantom{2}-1.07 \pm 0.1$ \\
$M^*_K (z=0)$  & $-22.26 $  (fixed) \\ 
$z_M$  & $\phantom{-2}1.78  \pm 0.15$\\
$k_M$  & $\phantom{-2}0.47 \pm  0.2$ \\
$z_{\phi}$ & $\phantom{-2}1.70 \pm 0.09$\\
$k_{\phi}$  & $\phantom{-2}1.47 \pm 0.1$\\ 
$\rm \phi_0 (z=0)$  & $\phantom{-}(3.5 \pm 0.4) \times 10^{-3}    \;\rm (Mpc^{-3})$ \\ \hline \hline 
\end{tabular}
\end{center}
\end{table}

First of all it is worth noting the excellent agreement between the estimation of the LF 
obtained with the two independent methods ($\rm 1/V_{max}$ and maximum likelihood) as shown in 
Fig. \ref{lf}. The large sample of galaxies provided by the UDS has allowed us to split the LF 
into narrow redshift and magnitude bins, while still preserving high statistical significance.

As a further check of the reliability of our results we find our estimate of the LF  to be
consistent with previous studies published in the literature.
In Fig. \ref{lf_data} we show the comparison with results obtained by 
Pozzetti et al. (2003) from the K20 survey, Caputi  et al. (2006) from 
the Great Observatories Origins Deep Survey (GOODS) / Chandra Deep Field
South (CDFS),  and Saracco et
al. (2006) from the Hubble deep field south (HDF-S). In particular, Saracco et al. exploited the 
very deep near-infrared observations of the HDF-S (FIRES, Franx et al. 2003) down to 
$K \le 24.9$ to derive an estimate of the faint end of the LF up to $z \sim 3$. They found
the value of the slope of the LF to be unchanged with redshift, consistent 
with the results we have obtained here from the maximum-likelihood analysis ($\alpha \simeq -1$). 
The small discrepancy in the normalization at $z \simeq 2$ between our LF and the one
estimated by Saracco et al. can be ascribed to the fact
that their LF is actually computed over a wider redshift range $1.9 \le z \le 4$, whereas 
we have simply plotted the LF at $z \simeq 2.5$.

Overall our estimate of the LF is in good agreement with previous results. However, the much
larger statistics provided by the UDS dataset has produced a substantial improvement in the
determination of the LF and its cosmological evolution. As clearly shown in Fig. \ref{lf}, the unique combination 
of area and depth provided by the UDS  allows the evolution of the LF, and in particular the bright end of the LF, 
to be accurately traced out to $z\geq3$. In fact, the
large increase in the areal coverage compared to previous studies -- more than a factor 16
compared to the CDFS and a factor of nearly 400 larger than the HUDF-FIRES -- has allowed us to trace the
evolution of rare, bright/massive galaxies with unprecedented accuracy.

\begin{figure}
\center{{
\epsfig{figure=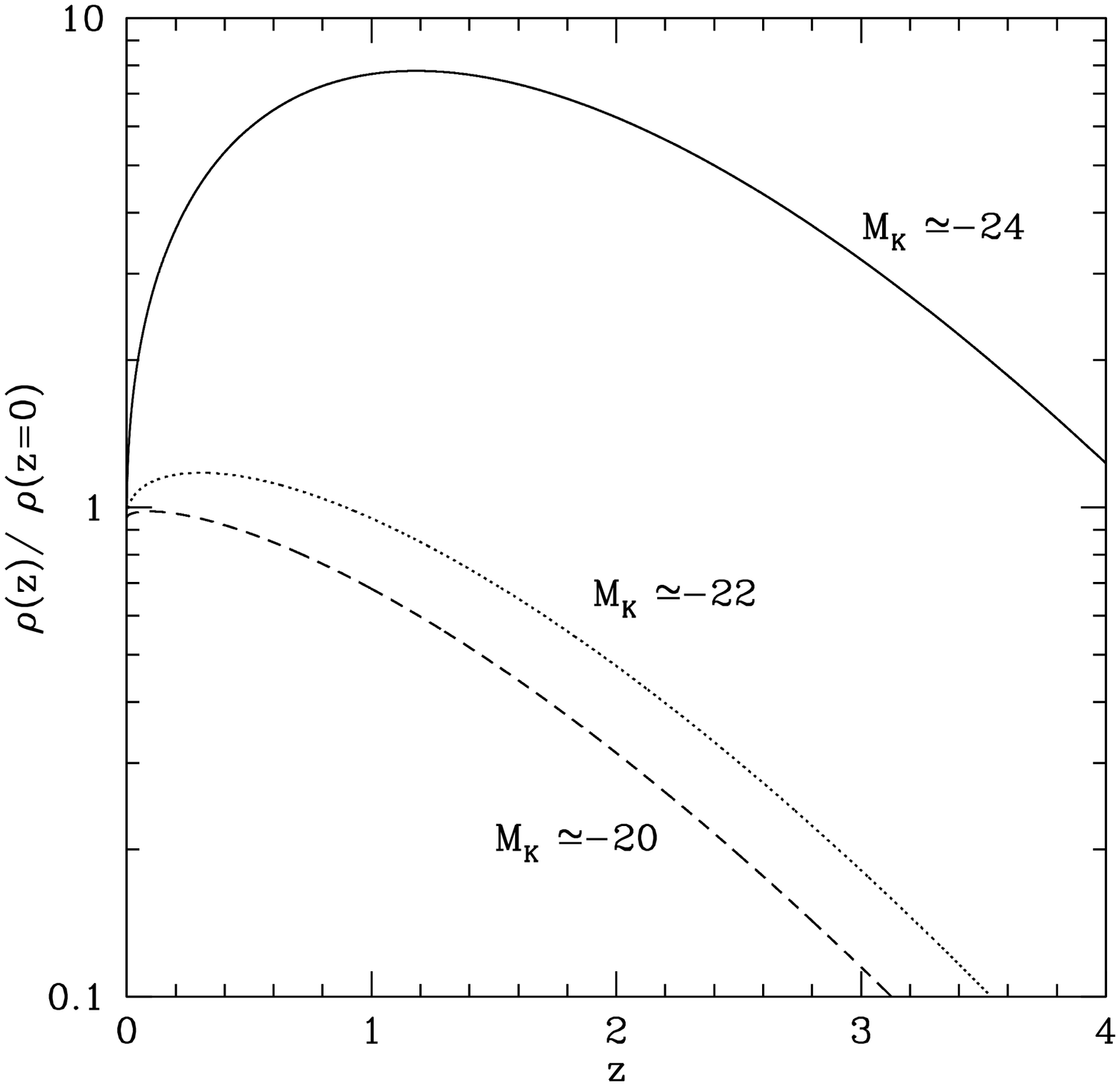, height=8cm}
\caption{\label{evol} Evolution of the space density of sources with different
rest-frame $K$-band luminosities as derived from the maximum-likelihood analysis and 
normalised to their corresponding local value.}
}}
\end{figure}

The maximum-likelihood analysis and the direct comparison of our results 
with the local $K$-band LF obtained by Kochanek et al. (2001) and Cole et al. (2001) suggest a
combination of luminosity and density evolution for the LF with cosmic time. In
agreement with previous studies (Pozzetti et al 2003; Drory et al. 2003; Feulner et al.
2003; Caputi et al. 2006; Saracco et al. 2006; Cirasuolo et al. 2007) we confirm the substantial
brightening of the characteristic luminosity by $\simeq1$ magnitude from $z =0 $ to $z \simeq 2$ and 
an overall decrement in the total normalization by a factor $\simeq 3.5$ over the same redshift interval. 
Given the exponential tail of the bright end of the LF the net result of
this combined luminosity and density evolution is that the space density of the brightest sources
with $M_K \simeq -24$ rapidly increases from the local Universe up to $z \simeq 1$ and then stays
roughly constant up to $z \simeq 2$, with a mild decline at higher redshift 
(see Fig. \ref{evol}).  This suggests that a large population of very bright/massive 
galaxies have assembled the bulk of their stars at high redshift, in the range $1 \simlt
z \simlt 3$. On the other hand the sharp decline in the space density of these bright systems at
$z \simlt 1$ is mainly due to the fading of the characteristic luminosity $M_K^*$,
consistent with the passive evolution of a single stellar population as shown in Cirasuolo et al.
(2007). This strongly suggests that, by $z \simeq 1$, massive galaxies have terminated their
mass assembly phase and that their luminosities then proceed to fade passively with cosmic time.

The evolutionary behaviour of less-luminous objects is somewhat different. As shown in Fig. \ref{evol} the
space density of galaxies with intermediate luminosity $M_K \simeq -22$  (the local value of the
characteristic luminosity $M_K^*$) is roughly constant up to $z \simeq 1$, 
but then decreases by more than a
factor of 5 by $z \simeq 3$. The evolution of less luminous galaxies (e.g. $M_K \simeq -20$) is even
more rapid, with a decline in the space density of a factor $\sim 2$ already apparent by $z \simeq 1$  and
by more than a factor 4 at $z \simeq 2$. It is worth noticing that in this latter case, with the current sample, we can
only directly trace the very faint sources with $M_K \simeq -20$ out to $z \sim 1.2$. At
higher redshift we rely on the extrapolation at low luminosities of the LF given by the maximum
likelihood analysis, but, as shown in Fig. \ref{lf_data}, the faint-end slope of the LF at
these redshifts is consistent with the results  obtained by Saracco et al. (2006).

\section{Comparison with predictions of theoretical models}
The presence of a large population of massive galaxies at high redshift has presented 
a major challenge for models of galaxy formation. 
In fact, models able to reproduce the local luminosity function
have often largely under-predicted the number of bright/massive galaxies at high redshift
(White \& Frenk 1991;  Kauffmann et al. 1993; Cole et al. 2000; Menci et al. 2002). 
On the other hand, models tuned to create more massive systems at high redshift 
have struggled to prevent further cooling flows and the continuous growth of these systems,
resulting in an over-prediction of massive systems in the local Universe 
(Kauffmann et al. 1999a,b).

The mutual feedback between star formation and accretion onto the central black hole has 
proven to be very effective in solving this problem. 
The energy injected into the system 
by nuclear activity is able to quench star-formation and expel the remaining cold gas from the 
galaxy (see Ciotti \& Ostriker 1997; Silk \& Rees 1998; Fabian 1999). This  
nuclear feedback  was first successfully introduced into a cosmological framework of galaxy formation by 
Granato et al. (2001, 2004) and then implemented by many other authors 
(e.g. Di Matteo et al. 2005; Croton et al. 2006; Bower et al. 2006; Menci et al 2006; 
Monaco et al. 2007).

In this section we compare our results with the predictions of some of 
the latest models of galaxy formation, most of them including some kind of recipe 
for active galactic nuclei (AGN) feedback.

Firstly, we considered two semi-analytical models by Bower et al. (2006) and De Lucia \& Blaizot 
(2007) which implement the physics of baryonic collapse, growth of the central black hole
and feedback from supernov\ae $\;$ and AGN on the Millennium simulation of the growth of the dark
matter structures in the $\Lambda$CDM cosmology (Springel et al. 2005).

Both models implement feedback from nuclear activity to quench cooling flows and regulate the 
formation of massive systems. Even though they utilise 
different prescriptions (described in Bower et al. 2006 and Croton et al. 2006),
both models assume the nuclear feedback 
to be driven by low-energy radio activity (the so-called ``radio mode'') which occurs
at low accretion rates and is found to be effective if the gas is in quasi hydrostic conditions, 
particularly at low redshift.

We also considered two other semi-analytical models which implement a rather different
treatment for the AGN feedback. Menci et al. (2006) present a model in which star-formation 
and BH growth is triggered by galaxy encounters, both merging and fly-by. In this model 
the nuclear feedback is produced during the short AGN phase,  where a blast wave from the 
central region is able to sweep the cold gas content of the galaxy on a shorter time 
scale and at higher redshift compared to the ``radio mode'' used by Bower et al. (2006) 
and Croton et al. (2006).

A similar ``quasar mode'' feedback, in which the radiation pressure from the nuclear 
activity blows out the cold gas from the galaxy has been adopted by Monaco et al. (2007) and 
Fontanot et al. (2008).
They have also implemented a revised treatment for the radiative cooling of the shocked 
gas and feedback from galactic winds and superwinds.

Finally, we compared our results with the predictions obtained with fully hydrodynamic 
simulations in a cosmological context by Nagamine et al. (2006) and Cen \& Ostriker (2006).
These simulations use a different approach from semi-analytical models, since they
solve a full set of hydro-dynamical equations for the evolution of dark matter and baryons 
simultaneously. These simulations implement the standard equations of gravity, hydrodynamics and
atomic physics, radiative transfer, heating and cooling of the gas and supernovae feedback.

\begin{figure*}
\center{{
\epsfig{figure=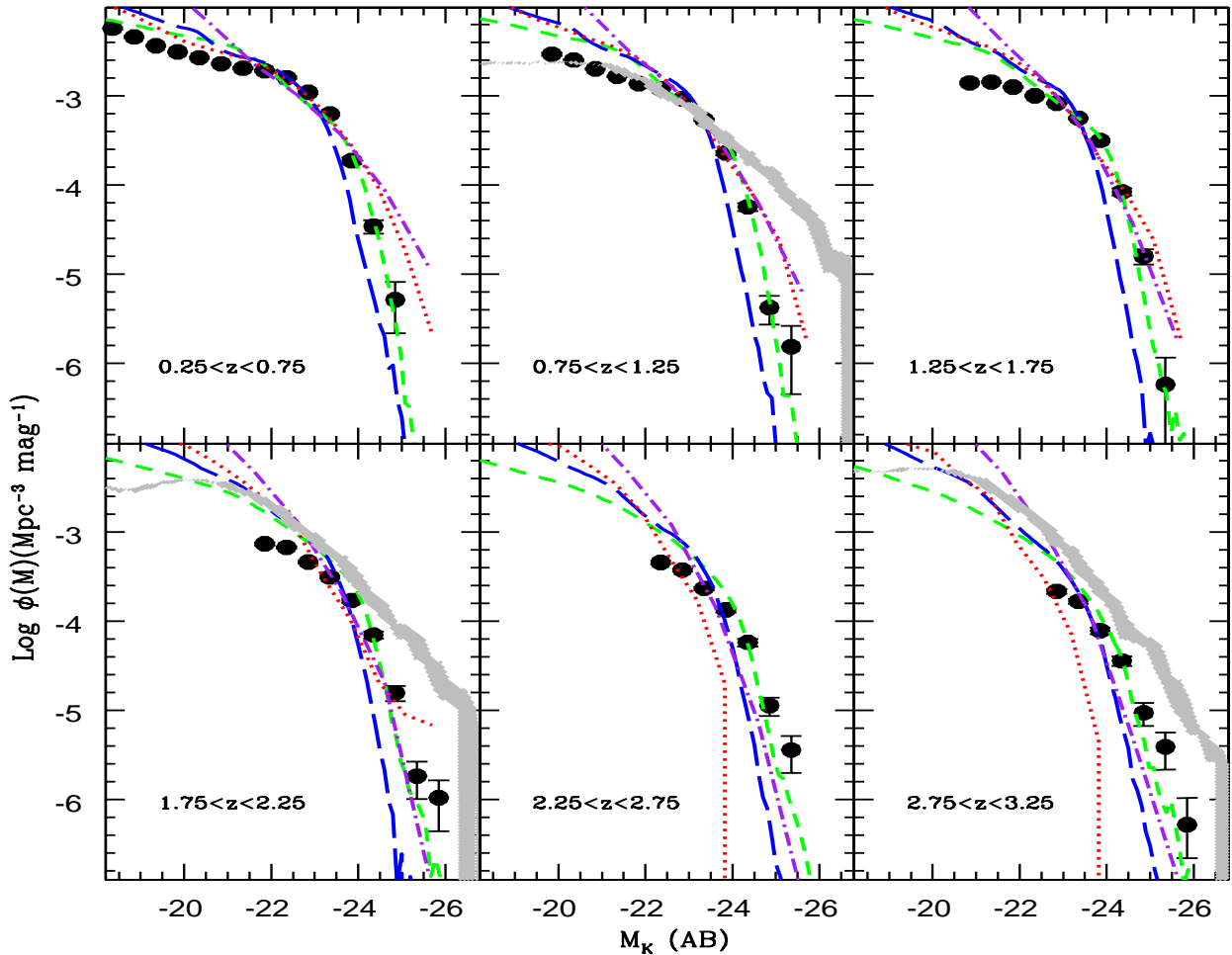, height=14cm, width= 18cm}
\caption{\label{lf_models} Comparison of our determination of the $K$-band LF (solid dots) with
predictions of theoretical models. Short and long dashed lines are the predictions obtained by
Bower et al. (2006) and De Lucia \& Blaizot (2007), respectively. The predictions by Monaco et
al. (2007) and Menci et al. (2006) are shown with a red dotted curve and purple dot-dashed curve, 
respectively.  The gray area shows the prediction obtained
by hydrodynamical simulations (Nagamine et al. 2006; Cen \& Ostriker 2006).}
}}
\end{figure*}

As shown in Fig. \ref{lf_models}, the new generation of theoretical models are better able
to reproduce the overall shape of the LF compared to previous models (Kauffmann et al.
1993; Cole et al. 2000; Menci et al. 2002).
In particular, the inclusion of the feedback from the central BH improves the quenching 
of star-formation in bright, massive systems, better reconciling the predicted bright end 
of the  LF with the observations. 

However, even though there is a broad consistency -- in particular with the 
prescriptions given by Bower et al. --  the large fraction of bright galaxies
already in place at high redshift is still challenging for the semi-analytical models. 
Especially at the highest redshifts, some of the  models still predict far fewer bright
galaxies than are observed. It is worth noticing that the predictions from  De Lucia \&
Blaizot (2007), Monaco et al (2007), Menci et al. (2006) and Nagamine et al. (2006) all include a correction for 
attenuation by dust, 
which is of course more realistic but introduces more uncertainties in the predicted LF, in
particular at the bright end.

On the other hand, the hydrodynamical simulations tend to over-predict the number of massive 
galaxies at any redshift. This could be due both to resolution effects and to a lack of 
feedback from the central AGN.
When the resolution is not sufficient, galaxies tend to over-merge in high-density regions.
The other issue is that the current 
hydrodynamical simulations do not yet include explicit implementation of AGN feedback, 
which has proven to be very effective in suppressing the bright end of the luminosity/mass
function. 

It is interesting to notice that most of the models are able to reproduce well the space 
density of galaxies at the knee of the
LF, but they consistently over-predict the space density of low-luminosity objects. 
This is probably due to very inefficient star formation in low-mass systems. 
In fact, from the modelling perspective, the feedback from AGN becomes negligible 
in these low-mass systems, and reducing the 
star-formation efficiency parameter also dramatically reduces the amount 
of feedback. This makes the discs gas rich and prone to low, but continuous, star-formation. This 
results in the over-prediction of the number of low-mass systems actually observed. 
Interestingly, the modelling of the star formation in the 
hydrodynamical simulation, at least around $z \simeq 1$, seems to be able to reproduce the shape
of the faint end of the LF, offering some promise that the problem of the excess of low-luminosity sources
can be resolved.

\section{Discussion and Conclusions}
We have presented new results on the cosmological evolution of the near-infrared galaxy
luminosity function from the local Universe up to $z \simeq 4$. The analysis is based on a large
and highly-complete sample of galaxies selected from the first data release of the UKIDSS Ultra
Deep Survey. Exploiting a master catalogue of $K$- and $z'$-selected galaxies 
over an area of 0.7 square degrees, we analysed a
sample of $\simeq 50,000$ galaxies,
all of which possess reliable photometry in 16 bands from the far-ultraviolet to the mid-infrared. This large
multi-wavelength coverage has allowed us to obtain accurate photometric redshifts and reliable
estimates of rest-frame $K$-band luminosities up to high redshift. 

The near-infrared LF and its cosmological evolution has been derived by using two independent
methods ($1/V_{max}$ and maximum likelihood) and the errors computed taking into account the
uncertainties in the photometric redshift estimation.
In particular, via the maximum-likelihood analysis we obtained a simple parameterization
for the LF and its cosmological evolution
which provides an excellent description of the data from $z =0$ up to $z \simeq 4$. 

The unique
combination of large area and depth offered by the UDS has 
allowed us to trace the evolution of the LF with
unprecedented accuracy. In agreement with previous studies, we find the evolution of the LF to be
well described by a combination of luminosity and density evolution. Given the shape of the LF
this results in a differential evolution of galaxy number density, dependent 
on galaxy luminosity, and reveals once again the downsizing behaviour of galaxy formation. 
Bright/massive galaxies are assembled at high redshift ($1 \simlt z \simlt 3$) and then they
passively evolved over the last few billion years, while the formation of intermediate and low
luminosity objects is progressively shifted to lower redshifts.

This anti-hierarchical behaviour is broadly reproduced by most of the latest generation of
theoretical models of galaxy formation. In this work we compared our observational results 
with the predictions of both semi-analytical models (Bower et al. 2006; De Lucia \& Blaizot 2007;
Monaco et al. 2007; Menci et al. 2006) and hydrodynamical simulations (Nagamine et al. 2006;
Cen \& Ostriker 2006). The inclusion of feedback from nuclear accretion has dramatically 
improved the ability of these models to reproduce the bright end of the LF. 
However, especially at high redshift, there is still a  wide scatter in the predicted number 
of luminous galaxies produced by the different  prescriptions for the physical processes 
implemented in the various semi-analytical models. 
On the other hand,
hydrodynamical simulations that have yet to incorporate AGN feedback, still systematically
over-predict the number of bright/massive galaxies. 

Although encouraging, these 
comparisons show that substantial discrepancies between the models and the data are still
present. 
In particular, all models tend to over-predict the number of low-luminosity galaxies at any
redshift. This is probably due to very low star-formation efficiency in real low-mass systems, 
which is difficult to implement in the models. This problem seems to be less severe in the
hydrodynamical simulations, but only at low redshift.

In conclusion, while the overall qualitative agreement between the data and the models is 
promising, some tension is still clearly present at both low and high luminosities. 
This suggests that probably some fundamental physical process(es) is(are) still missing in the models, 
or that the models need to be further fine tuned. This also shows how detailed and accurate observational estimates of
the space density of galaxies, in particular at high redshift, can prove to be a very 
powerful tool for the testing of galaxy-formation models. 
In the near future we aim to exploit the forthcoming deeper {\it Spitzer} IRAC
and MIPS observations in the UDS to extend the study presented here to the 
direct derivation of stellar-mass growth and star-formation as a function of 
cosmic time.

\section{acknowledgements}
MC, SF and CS would like to acknowledge funding from
STFC.
RJM, OA would like to acknowledge the funding of the Royal
Society.  We are grateful to R.C. Bower, K. Nagamine, P. Monaco, F. Fontanot, 
G. De Lucia and N. Menci for having provided us with the predictions from their 
latest models.
We are also grateful to the staff at UKIRT and Subaru for making these 
observations possible. We also acknowledge the Cambridge Astronomical
Survey Unit and the Wide Field Astronomy Unit in Edinburgh for processing the 
UKIDSS data.

\end{document}